\documentclass[10pt,twocolumn,letterpaper]{article}

\usepackage{iccv}
\usepackage{times}
\usepackage{epsfig}
\usepackage{graphicx}
\usepackage{amsmath}
\usepackage{amssymb}

\usepackage{cite}
\usepackage{amsmath,amssymb,amsfonts}
\usepackage{algorithmic}
\usepackage{graphicx}
\usepackage{textcomp}
\usepackage{xcolor}

\usepackage{url}
\usepackage{algorithm,algorithmic}
\usepackage{booktabs,makecell,multirow}

\usepackage[breaklinks=true,bookmarks=false]{hyperref}

\iccvfinalcopy 

\def\iccvPaperID{12} 

\ificcvfinal\fi

\begin{document}

\title{Continuous Emotion Recognition using Visual-audio-linguistic Information: A Technical Report for ABAW3}

\author{Su Zhang\textsuperscript{1}, Ruyi An\textsuperscript{1}, Yi Ding\textsuperscript{1}, Cuntai Guan\textsuperscript{1, \thanks{This work is partially supported by the RIE2020 AME Programmatic Fund, Singapore (No. A20G8b0102).}} \\
\textsuperscript{1}Nanyang Technological University\\
{\tt\small sorazcn@gmail.com, ran003@e.ntu.edu.sg, ding.yi@ntu.edu.sg, ctguan@ntu.edu.sg}
}

\maketitle
\ificcvfinal\thispagestyle{empty}\fi

\begin{abstract}
We propose a cross-modal co-attention model for continuous emotion recognition using visual-audio-linguistic information. The model consists of four blocks. The visual, audio, and linguistic blocks are used to learn the spatial-temporal features of the multi-modal input. A co-attention block is designed to fuse the learned features with the multi-head co-attention mechanism. The visual encoding from the visual block is concatenated with the attention feature to emphasize the visual information. To make full use of the data and alleviate over-fitting, cross-validation is carried out on the training and validation set. The concordance correlation coefficient (CCC) centering is used to merge the results from each fold. The achieved CCC on the test set is $0.520$ for valence and $0.602$ for arousal, which significantly outperforms the baseline method with the corresponding CCC of $0.180$ and $0.170$ for valence and arousal, respectively. The code is available at https://github.com/sucv/ABAW3.
\end{abstract}

\section{Introduction}
Emotion recognition is the process of identifying human emotion. It plays a crucial role in behavioral modeling, human-computer interaction, and affective computing. By using the dimensional model \cite{sandbach2012static}, any emotional state can be taken as a point located in a continuous space, with the valence and arousal being the axes. Continuous emotion recognition seeks to map the $N$ sequential data points into $M$ sequential emotional state points, where $M$ usually equals $N$. This report details our methodology for the valence-arousal estimation challenge from the third affective behavior analysis in-the-wild (ABAW3) workshop \cite{kollias2022abaw, kollias2021analysing, kollias2020analysing, kollias2021distribution, kollias2021affect, kollias2019expression, kollias2019face, kollias2019deep, zafeiriou2017aff}.

Our work is an extension of the last year's attempt \cite{zhang2021continuous} on ABAW2 \cite{kollias2021analysing}. The audio-visual-linguistic information is exploited and combined by using a multi-head attentive feature fusion scheme. The network consists of three branches, fed by synchronous video frames, VGGish \cite{hershey2017cnn} features, and BERT features. The visual branch consists of a Resnet50 for spatial encoding and a temporal convolutional network (TCN) \cite{bai2018empirical} for temporal encoding. The audio and linguistic branches each contain a TCN for temporal encoding. The three branches work in parallel and their outputs are sent to the attentive fusion block. The latter employs the co-attention mechanism of the three branches for feature fusion. To emphasize the dominance of the visual features, which we believe to have the strongest correlation with the label, the temporal feature outputted by the TCN of the visual branch is concatenated to the fused feature. A fully-connected layer is used for the regression. The 6-fold cross-validation is carried out to alleviate the over-fitting. 


\begin{figure*}[t]
\centering
\includegraphics[width=\textwidth]{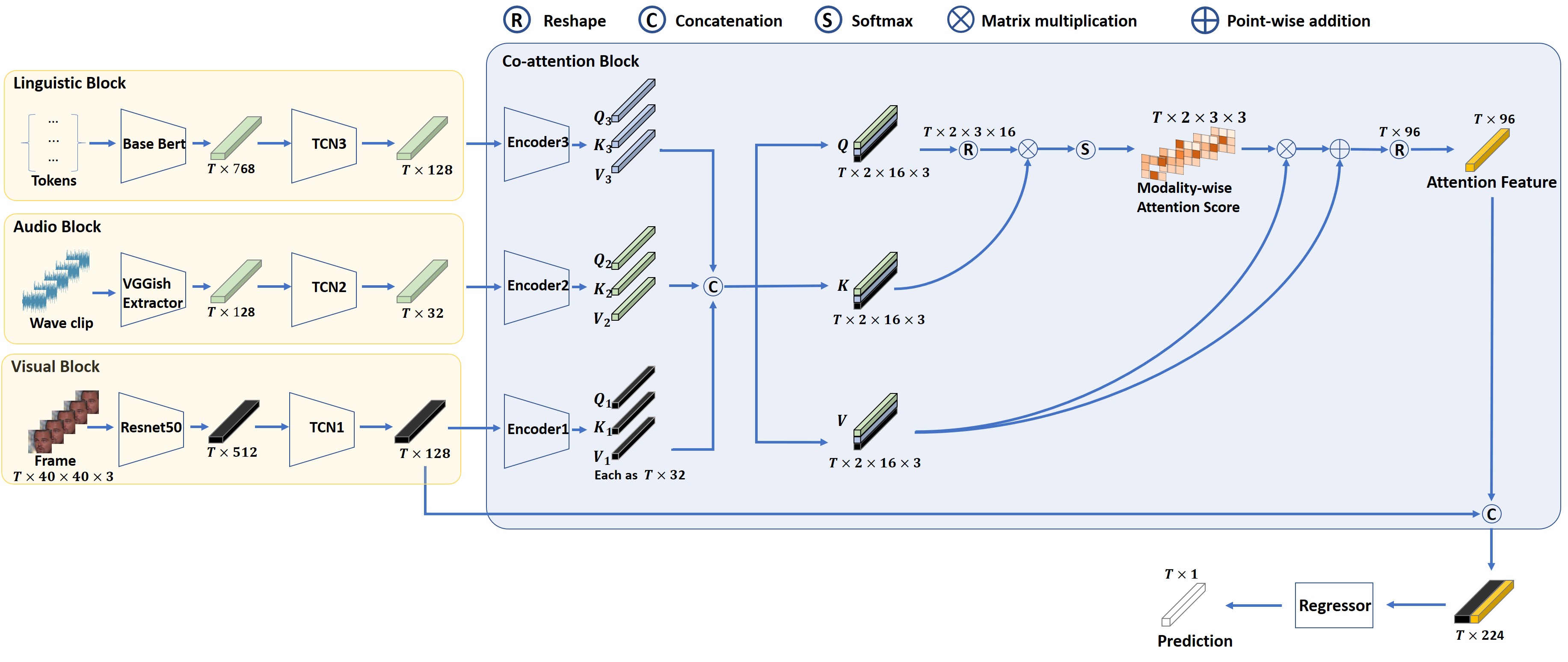}
\caption{The architecture of our proposed model. The model consists of four components, i.e., the visual, audio, linguistic, and co-attention blocks. The visual block has a cascaded 2DCNN-TCN structure, and the audio and linguistic blocks each contain a TCN. The three branches yield three independent spatial-temporal feature vectors. They are then fed to the attentive fusion block. Three independent attention encoders are used. For the $i$-th branch, its encoder consists of three independent linear layers, they adjust the dimension of the feature vector producing a query $\mathbf{Q}_i$, a key $\mathbf{K}_i$, and a value $\mathbf{V}_i$. They are then regrouped and concatenated to form the cross-modal counterparts. For example, the cross-modal query $\mathbf{Q}=[\mathbf{Q}_1,\mathbf{Q}_2,\mathbf{Q}_3]$. An attention score is obtained by Eq. \ref{eq:att}.}\label{fig:arc}
\end{figure*}

The remainder of the paper is arranged as follows. Section \ref{sec:model_architecture} details the model architecture including the visual, aural, linguistic, and attentive fusion blocks. Section \ref{sec:implementation_details} elaborates the implementation details including the data pre-processing, training settings, and post-processing. Section \ref{sec:result} provides the continuous emotion recognition results on the Aff-Wild2 database. Section \ref{sec:conclusion} concludes the work.

\section{Model Architecture}
\label{sec:model_architecture}
The model architecture is illustrated in Fig. \ref{fig:arc}. The visual block consists of a Resnet50 and a TCN. The resnet50 plays the role of backbone and is pre-trained on the MS-CELEB-1M dataset as a facial recognition task, it is then fine-tuned on the FER+ \cite{barsoum2016training} dataset. The Resnet50 spatially encodes each frame from a video frame sequence. The sequential spatial encodings are then stacked and fed to a TCN for temporal encoding. The audio and linguistic blocks each consist of a TCN. The Vggish \cite{hershey2017cnn} and BERT \cite{radford2019language} features are the inputs, respectively. 

Given the three temporal encodings from the visual, audio, and linguistic blocks, the attentive fusion block first maps the feature vectors to query, key, and value vectors by the following procedure. For the $i$-th branch, its encoder consists of three independent linear layers, they adjust the dimension of the feature vector producing a query $\mathbf{Q}_i$, a key $\mathbf{K}_i$, and a value $\mathbf{V}_i$. They are then regrouped and concatenated to form the cross-modal counterparts. For example, the cross-modal query $\mathbf{Q}=[\mathbf{Q}_1,\mathbf{Q}_2,\mathbf{Q}_3]$.  After which, the attention feature is calculated as:

\begin{equation}
Attention(\mathbf{Q}, \mathbf{K}, \mathbf{V})=(softmax(\frac{\mathbf{QK}^T}{\sqrt{d_K}})+1)\mathbf{V},
\label{eq:att}
\end{equation}
where $d_K=32$ is the dimension of the key $\mathbf{K}$. After which, the attention feature is normalized and concatenated to the visual temporal encoding. Finally, a fully connected layer is employed to infer. Following the labeling scheme of the Aff-Wild2 database where the label frequency equals the video frame rate, each frame of the input video sequence is exactly corresponding to one label point.

\section{Implementation Details}
\label{sec:implementation_details}
\subsection{Database}
The ABAW3 competition uses the Aff-Wild2 database. The corpora of the valence-arousal estimation sub-challenge includes $564$ trials. The database is split into the training, validation and test sets. The partitioning is done in a subject independent manner so that any subject's data are included in only one partition. The partitioning produces $341$, $71$, and $152$ trials for the training, validation, and test sets. Four experts annotate the videos using the method proposed in \cite{cowie2000feeltrace}. In addition to the annotations (for the training and validation sets only) and the raw videos, the bounding boxes and landmarks for each raw video are also available to the participants.

\subsection{Preprocessing}
The visual preprocessing is carried out as follows. The cropped-aligned image data provided by the organizer are used. All the images are resized to $48\times 48\times 3$. Given a trial from the training or validation set, the length $N$ is determined by the number of the rows of the annotation text file which does not include $-5$. For the test set, the length $N$ is determined by the frame number of the raw video. A zero matrix $\mathbf{B}$ of size $N\times 48\times 48\times 3$ is initialized and then iterated over the rows. For the $i$-th row of $\mathbf{B}$, it is assigned as the $i$-th jpg image if it exists, otherwise doing nothing. 

The audio preprocessing firstly converts all the videos to mono with a $16K$ sampling rate in wav format. The VGGish features are then extracted using the pretrained Vggish model\footnote{https://github.com/harritaylor/torchvggish}. The only change is that we specified the hop length to be $1/frame rate$ of the raw video, in order to synchronize with other modalities and annotations.  

The linguistic preprocessing is carried out as follows. The mono wav file obtained from the audio preprocessing is fed to a pretrained speech recognition model from the Vosk toolkit\footnote{https://alphacephei.com/vosk/models/vosk-model-en-us-0.22.zip}, from which the recognized words and the word-level timestamp are obtained. The recognized words are then fed to a pretrained punctuation and capitalization model from the Nvidia Nemo toolkit\footnote{https://docs.nvidia.com/deeplearning/nemo/user-guide/docs/en/main/nlp/punctuation\_and\_capitalization.html}. After which a pretrained BERT model from the Pytorch library is employed to extract the word-level linguistic features. The linguistic features are obtained by summing together the last four layers of the BERT model \cite{sun2020multi}. To synchronize, the word-level linguistic features are populated according to the timestamp of each word and each frame. Specifically, a word usually has a larger time span than that for a frame. Therefore, for one word, its feature is repetitively assigned to the time steps of all the frames within the time span.

For the valence-arousal labels, all the rows containing $-5$ are excluded. To ensure that the features have the same length as the corresponding trial, the feature matrices are either repeatedly padded (using the last feature points) or trimmed (starting from the rear), depending on whether the feature length is shorter than the trial length or not, respectively.

\subsection{Data Expansion}

The AffWild2 database employed by ABAW3 contains $341$ and $71$ trials in the training and validation sets, respectively. To clarify, a label txt file and its corresponding data are taken as a trial. Note that some videos include two subjects, resulting in two separated cropped-aligned image folders and label txt files, with different suffixes. They are each taken as two trials. Note that there are $10$ fewer trials for the training set compared to its counterpart from ABAW2.

To make full use of the available data and alleviate over-fitting, 6-fold cross-validation is employed. By evenly splitting the training set into $5$ folds, we have $6$ folds in total with a roughly equal trial amount, i.e., $68\times 4+69+71$ trials. Note that the $0$-th fold is exactly the original data partitioning. And there is no subject overlap across different folds. The CCC-centering is employed to merge the inference result on the test set.

Moreover, during training and validation, the resampling window has a $33\%$ overlap, resulting in $33\%$ more data. 

\subsection{Training}
The batch size is $2$. For each batch, the resampling window length and hop length are $300$ and $200$, respectively. I.e., the dataloader loads consecutive $300$ feature points to form a minibatch, with a stride of $200$. For any trials having feature points smaller than the window length, zero padding is employed. For visual data, the random flip, random crop with a size of $40$ are employed for training and only the center crop is employed for validation. The data are then normalized so that $mean=std=0.5$. For audio and linguistic data, they are normalized so that $mean=0$ and $std=1$.

The CCC loss is used as the loss function. The Adam optimizer with a weight decay of $0.001$ is employed. The learning rate (LR) and minimal learning rate (MLR) are set to $1e-5$ and $1e-7$, respectively. The \textit{ReduceLROnPlateau} scheduler with a patience of $5$ and factor of $0.1$ is employed based on the validation CCC. The maximal epoch number and early stopping counter are set to $100$ and $10$, respectively. Two groups of layers for the Resnet50 backbone are manually selected for further fine-tuning, which corresponds to the whole layer4 and the last three blocks of layer3. 

\begin{table*}[ht]
\centering
\caption{The CCC results from the 6-fold cross-validation on the validation and test sets. Fold 0 is exactly the original data partitioning provided by ABAW3. CCCC denotes CCC-centering. Since 5 submissions are allowed, there are no test results on Fold 2 and 3.}\label{table:result}
\setlength{\tabcolsep}{4mm}{
\begin{tabular}{cccccccccc}
\toprule
 
\makecell[c]{Emotion} &\makecell[c]{Partition}&\makecell[c]{Method}&\makecell[c]{ Fold 0}& \makecell[c] {{\textbf{Fold 1}}}  &\makecell[c]{Fold 2} &\makecell[c]{Fold 3} &\makecell[c]{Fold 4} &\makecell[c]{ Fold 5} &\makecell[c]{CCCC}         \\
 
\midrule
\multirowcell{4}{Valence}&\multirowcell{2}{Validation}&Baseline&$0.310$&$-$&$-$&$-$&$-$&$-$&$-$\\
&&Ours&$0.450$&$\mathbf{0.559}$&$0.469$&$0.531$&$0.539$&$0.448$&$-$\\
\cmidrule(r){3-10}
&\multirowcell{2}{Test}&Baseline&$0.180$&$-$&$-$&$-$&$-$&$-$&$-$\\
&&Ours&$0.490$&$\mathbf{0.520}$&$-$&$-$&$0.479$&$0.511$&$0.490$\\
\midrule
\multirowcell{4}{Arousal}&\multirowcell{2}{Validation}&Baseline&$0.170$&$-$&$-$&$-$&$-$&$-$&$-$\\
&&Ours&$0.651$&$\mathbf{0.671}$&$0.564$&$0.562$&$0.631$&$0.618$&$-$\\
\cmidrule(r){3-10}
&\multirowcell{2}{Test}&Baseline&$0.170$&$-$&$-$&$-$&$-$&$-$&$-$\\
&&Ours&$0.584$&$\mathbf{0.602}$&$-$&$-$&$0.580$&$0.587$&$0.573$\\

\bottomrule
\end{tabular}}
\end{table*}

The training strategy aims to alleviate the out-fitting issue. The Resnet50 backbone is initially fixed except for the output layer. The learning rate is linearly warmed up to $1e-5$ when epoch $\leq$ 10. After that, the PyTorch \textit{ReduceLROnPlateau} steps in as the scheduler. The CCC on the validation set determines whether the training improves. If there were no improvements for 5 epochs, the scheduler would reduce the $LR$ to $0.1\cdot LR$. When $LR<1e-7$, layer4 of the backbone is unfrozen and the scheduler is reset. This process repeats until the last three blocks of layer3 are unfrozen. Finally, when once again $LR<1e-7$, the training is done. Note that at the end of each epoch, the current best model state dictionary (i.e., the one with the greatest validation CCC) is loaded. By doing so, when there is no improvement on validation CCC, the training CCC cannot keep increasing and ends up with over-fitting.

\subsection{Post-processing}

The post-processing consists of CCC-centering and clipping. 
Given the predictions from 6-fold cross-validation, the CCC-centering aims to yield the weighted prediction based on the inter-class correlation coefficient (ICC) \cite{ringeval2019avec}. This technique has been widely used in many emotion recognition challenges \cite{valstar2016avec,ringeval2017avec,ringeval2018avec,ringeval2019avec} to obtain the gold-standard labels from multiple raters, by which the bias and inconsistency among individual raters are compensated. The clipping ensures that the inference is truncated within the interval $[-1, 1]$, i,e., any values larger or smaller than $1$ or $-1$ are set to $1$ or $-1$. respectively. 

With all the predictions from the 6-fold cross-validation and CCC-centering, we have 5 submissions allowed. They are determined as follows. On the one hand, based on the experience from ABAW2, we find that the influence of the CCC-centering and clipping on the test CCC could be quite trivial ($\pm 0.001$). On the other hand, we also tried other centering methods, such as the estimator weighted evaluator \cite{schuller2013intelligent} and rater aligned annotation weighting \cite{stappen2021muse}. However, there is almost no visual difference among the three centering methods when we visualize the centered results trial by trial. Finally, our five submissions consist of the predictions from the CCC-centering, Fold0, and the three best folds.

\section{Result}
\label{sec:result}

The validation and test results of our method against the baseline are reported in Table \ref{table:result}. The best results from all the teams are reported in Table \ref{table:all}.

\begin{table}[ht]
\centering
\caption{The overall test results in CCC. The bold fonts indicate the best results. Citations for several teams are not available from Google Scholar by the time when this report was drafted. }\label{table:all}
\vspace{0.25cm}
\begin{tabular}{|p{3cm}|l|l|l|}
\hline
Method                     & Valence & Arousal & Mean  \\ \hline
Situ-RUCAIM3 \cite{meng2022multi} & \textbf{0.606} & 0.596 & \textbf{0.601}\\ \hline
Ours  & 0.520 & \textbf{0.602} & 0.561 \\ \hline
PRL \cite{nguyen2022ensemble} & 0.450 & 0.445 & 0.448 \\ \hline
HSE-NN  & 0.417 & 0.454 & 0.436 \\ \hline
AU-NO  & 0.418 & 0.407 & 0.413 \\ \hline
LIVIA-2022  & 0.374 & 0.363 & 0.369 \\ \hline
Netease Fuxi Virtual Human \cite{zhang2022transformer} & 0.300 & 0.244 & 0.272 \\ \hline
Baseline \cite{kollias2022abaw}                  & 0.180             & 0.170             & 0.175  \\ \hline 
\end{tabular}
\end{table}



\section{Conclusion}
We propose a cross-modal co-attention model for continuous emotion recognition using visual-audio-linguistic information. The model employs three parallel blocks to learn the feature from the three modalities. The learned features are then fused by the cross-modal co-attention block. Experiments are conducted on the Aff-Wild2 database. The achieved CCC on the test set is $0.520$ for valence and $0.602$ for arousal, which significantly outperforms the baseline method with the corresponding CCC of $0.180$ and $0.170$ for valence and arousal, respectively.

\label{sec:conclusion}
\bibliographystyle{IEEEtran}
\bibliography{ref}

\end{document}